\documentclass[journal]{IEEEtran}

\usepackage{verbatim}
\usepackage{amsfonts}
\usepackage{amssymb}
\usepackage{stfloats}
\usepackage{cite}
\usepackage{graphicx}
\usepackage{psfrag}
\usepackage{amsmath}
\usepackage{array}
\usepackage{epstopdf}
\usepackage{authblk}
\usepackage{graphicx}
\usepackage{amsthm}
\usepackage{lipsum}
\usepackage{verbatim}
\usepackage{authblk}
\usepackage{mathtools}
\usepackage{cuted}
\usepackage{booktabs}

\usepackage{amsmath}
\usepackage{mathrsfs}

\usepackage{algorithmic}

\usepackage{array}

\ifCLASSOPTIONcompsoc
\usepackage[caption=false,font=normalsize,labelfont=sf,textfont=sf]{subfig}
\else
\usepackage[caption=false,font=footnotesize]{subfig}
\fi

\usepackage{url}
\usepackage{graphicx,amsmath,amssymb,amsfonts}
\usepackage{algorithmic,algorithm}

\usepackage{hyperref}
\hypersetup{colorlinks=true, citecolor=blue}

\usepackage{setspace}

\makeatletter

\newcommand{\Rmnum}[1]{\expandafter\@slowromancap\romannumeral #1@}
\makeatother

\usepackage{framed}

\usepackage{cancel}

\usepackage{booktabs}
\usepackage{tabularx,makecell,multirow}

\usepackage[table]{xcolor}
\usepackage{pifont}

\usepackage{diagbox}

\definecolor{shadecolor}{rgb}{0.92,0.92,0.92}

\begin{document}
	\bstctlcite{ref:BSTcontrol}

	\title{\huge Direct Satellite-to-Device Communications: From Cooperative Task Offloading to Non-Cooperative Access Monitoring}
	
	\author{Sai Huang, Wanli Ni, Ke Lv, Pengcheng Zhang, Yurui Zheng, Menghan Zhang, Zihui Gong, and Zhiyong Feng
		
		\thanks{This work is supportted by National Natural Science Foundation of China under Grant 62595742, 62422103, 62321001, 62501351, and Fundamental Research Funds for the Central Universities under Grant 2025U40-2. (Corresponding author: Wanli Ni)}
		\thanks{All authors are with the Key Laboratory of Universal Wireless Communications, Ministry of Education, Beijing University of Posts and Telecommunications, Beijing 100876, China.} 
	}
	
	\maketitle

	\begin{abstract}
		Direct satellite-to-device (DS2D) communication is emerging as a transformative paradigm for extending ubiquitous connectivity and edge computing capabilities to remote and underserved regions within 6G non-terrestrial networks. However, practical deployment faces dual critical challenges: {\color{black}i) dynamic satellite channel conditions (e.g., severe Doppler shifts, fast fading) and constrained satellite computing resources in cooperative scenarios; and ii) unauthorized satellite access introduces significant spectrum security threats in non-cooperative scenarios.}
		To address these challenges, we propose a versatile DS2D system that supports cooperative task offloading and non-cooperative access monitoring.
		For cooperative DS2D communications, we integrate a channel estimation module with a dueling double deep Q-network (D3QN) to dynamically optimize task offloading strategy.
		For non-cooperative DS2D communications, we propose Transformer-based models to enable \textcolor{black}{blind signal detection} and automatic modulation classification (AMC).
		Simulation results show that:
		1) The D3QN algorithm reduces average latency by up to 225\% compared to static association policies.
		2) Our signal detection model achieves an average presence detection probability of 90.5\% for DS2D signals.
		3) The proposed AMC algorithm achieves superior performance across different signal-to-noise ratios (SNRs), with a 9.4\% accuracy gain in low-SNR regimes compared to existing methods.
	\end{abstract}
	
	\vspace{-2 mm}
	\section{Introduction}
	Terrestrial cellular networks, despite their high capacity and low latency, are inherently constrained by geographical limitations, high infrastructure costs, and susceptibility to natural disasters.
	The rapid proliferation of mobile broadband and mission-critical Internet of Things (IoT) applications has intensified the demand for \textcolor{black}{ubiquitous connectivity}, particularly in remote regions or disaster-stricken areas where terrestrial base stations may be damaged or absent.
	In response, direct satellite-to-device (DS2D) communication has emerged as a transformative solution, enabling low Earth orbit (LEO) satellite constellations to establish direct links with unmodified terrestrial user equipment (UE), including commercial smartphones, handsets, and IoT devices \cite{Bakhsh2025Survey}.
	\textcolor{black}{Consequently, by delivering reliable wide-area connectivity without reliance on ground relays and enabling space-ground computing for remote and underserved users, DS2D is becoming a cornerstone of 6G integrated satellite–terrestrial networks \cite{An2024C}.}
	\textcolor{black}{Beyond merely filling coverage gaps, DS2D is now evolving into an operator-grade networking capability that can support seamless global services for both consumer and critical applications.}

	Early DS2D deployments primarily target narrowband messaging and emergency services, while the long-term vision encompasses broadband access, real-time sensing, and edge intelligence \cite{An2024C, Wei2025Network}. However, the user experience of DS2D services faces significant challenges due to weak handheld-to-satellite links, severe Doppler shifts, and intermittent satellite visibility \cite{Li2026T}.
	\textcolor{black}{Beyond technical limitations, DS2D communications introduce critical security concerns. Unauthorized users may exploit non-cooperative or unregulated satellites to establish unsupervised links. Monitoring such communications presents formidable challenges, as receivers typically lack prior knowledge of essential signal parameters.
	Compounded by low signal-to-noise ratio (SNR) reception and Doppler-induced distortions, these factors render blind signal detection and modulation classification exceptionally difficult.}

	Recent research has made significant progress in addressing the challenges of cooperative or non-cooperative DS2D communications, though largely in isolation.
	On the cooperative front, Zheng \textit{et al.} \cite{Zheng2024D} proposed a data- and model-dual-driven deep unfolding network for joint activity detection and channel estimation in multibeam LEO systems, substantially reducing pilot overhead under grant-free random access.
	Furthermore, He \textit{et al.} \cite{He2026C} developed a collaborative channel estimation and prediction framework that mitigates channel aging by integrating sparse Bayesian learning with long short-term memory (LSTM)-based forecasting.
	In the context of integrated space-ground computing, Peng \textit{et al.} \cite{Peng2025I} investigated intelligent task offloading in satellite–cloud–edge architectures for vehicular networks, employing deep reinforcement learning (DRL) to dynamically orchestrate workloads across orbital and terrestrial tiers.
	Meanwhile, Li \textit{et al.} \cite{Li2026T} introduced a time-synchronization-aided signal detection scheme that narrows the Doppler and delay search ranges, thereby enhancing acquisition reliability in LEO downlinks.
	Turning to non-cooperative scenarios, Uvaydov \textit{et al.} \cite{Uvaydov2024S} pioneered wideband semantic spectrum segmentation using an enhanced U-Net architecture, enabling pixel-level identification of unknown communication modes.
	For modulation recognition under adverse conditions, An \textit{et al.} \cite{An2024C} designed a Transformer-based classifier coupled with blind equalization to combat inter-beam interference and imperfect channel state information (CSI) in multibeam satellite systems.
	Shao \textit{et al.} \cite{Shao2025IQ} proposed IQFormer, a multi-modality fusion Transformer that jointly processes in-phase/quadrature (I/Q) and time-frequency representations to achieve robust automatic modulation classification (AMC) even at low SNRs.

	As compared in Table \ref{table}, distinct from existing approaches, in this article, we propose a unified DS2D framework that supports cooperative task offloading and non-cooperative access monitoring, aiming to provide a holistic foundation for trustworthy, scalable, and intelligent DS2D ecosystems in the 6G era.
	Specifically, the main contributions of this work are summarized as follows:
	\begin{itemize}
		\item
		For cooperative computing resource orchestration, we propose a DRL algorithm that integrates channel estimation with adaptive Doppler compensation and user association decisions, achieving up to 225\% latency reduction compared to static association policies.
		\item
		For non-cooperative DS2D scenarios, we first introduce a segment anything model (SAM)-based signal detection method that achieves a detection probability exceeding 88\% while keeping the false alarm rate below 4.5\%. Then, we propose a CNN-Transformer-based dual-stream model for AMC, which attains 91.6\% recognition accuracy in high-SNR regimes and maintains 70.5\% accuracy under low-SNR conditions.
	\end{itemize}

	\begin{table*}[htbp]
		\centering
		\caption{Comparison of this article with related work. Abbreviations: CE, RA, SD, and AMC denote channel estimation, resource allocation, signal detection, and automatic modulation classification, respectively; OTFS, LSTM, and CNN stand for orthogonal time frequency space, long short-term memory, and convolutional neural network.}
		\label{table}
		\begin{tabular}{@{}
				>{\centering\arraybackslash}p{0.6cm}   
				>{\raggedright\arraybackslash}m{3.3cm}   
				>{\centering\arraybackslash}m{0.3cm}     
				>{\centering\arraybackslash}m{0.3cm}     
				>{\centering\arraybackslash}m{0.3cm}     
				>{\centering\arraybackslash}m{0.3cm}     
				>{\raggedright\arraybackslash}m{4.3cm}   
				>{\raggedright\arraybackslash}m{4.6cm}   
				@{}}
			\toprule
			{\textbf{Ref.}} & {\textbf{Scenario}} & \multicolumn{1}{c}{\textbf{CE}} & {\textbf{RA}} & {\textbf{SD}} & {\textbf{AMC}} & \multicolumn{1}{c}{\textbf{Methodology}} & \multicolumn{1}{c}{\textbf{Considerations}} \\
			\midrule
			\cite{Zheng2024D} & Cooperative DS2D & \checkmark & \ding{55} & \ding{55} & \ding{55} & Model-driven deep unfolding & Metasurface-aided LEO,  grant-free random access \\
			
			\cite{He2026C} & Cooperative DS2D & \checkmark & \ding{55} & \ding{55} & \ding{55} & Sparse Bayesian and LSTM  & OTFS modulation, downlink channel aging \\ 
			
			\cite{Peng2025I} & Cooperative DS2D  & \ding{55} & \checkmark & \ding{55} & \ding{55} & Deep reinforcement learning & LEO-aided offloading, dynamic task arrivals \\
			
			\cite{Li2026T} & Cooperative DS2D  & \ding{55} & \ding{55} & \checkmark & \ding{55} & Time synchronization-aided acquisition & Low SNR, known user location, signal monitoring \\
			
			\cite{An2024C} & Non-cooperative DS2D & \ding{55} & \ding{55} & \ding{55} & \checkmark  & Blind equalization, CNN-Transformer model & Inter-beam interference, imperfect CSI \\
			
			\cite{Uvaydov2024S} & Non-cooperative communication & \ding{55} & \ding{55} & \checkmark & \ding{55} & Collaborative learning, improved U-Net &  Communication mode identification \\
			
			\cite{Shao2025IQ} & Non-cooperative communication & \ding{55} & \ding{55} & \ding{55} & \checkmark & Transformer-based model, multi-modal fusion & Low SNR, unknown modulation type \\
			
			Ours & Cooperative \& non-cooperative DS2D & \checkmark & \checkmark & \checkmark & \checkmark & Deep reinforcement learning, Transformer-based model &  Satellite edge computing, unauthorized access monitoring \\
			\bottomrule
		\end{tabular}
	\end{table*}

	\begin{figure*}[t]
		\centering
		\includegraphics[width=6.3 in]{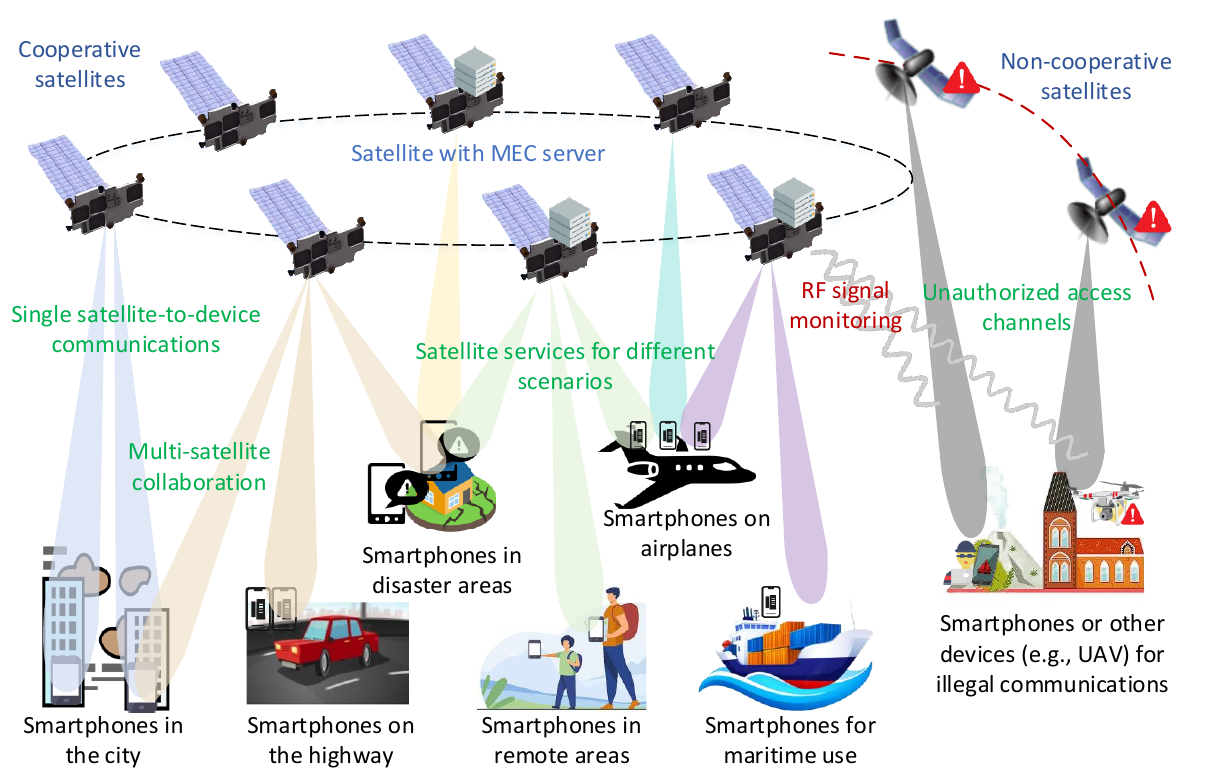}
		\caption{An illustration of DS2D communications for both cooperative and non-cooperative satellite scenarios.
		On the left, a constellation of cooperative satellites enables seamless connectivity across diverse terrestrial environments, including urban areas, highways, remote regions, disaster zones, maritime operations, and airborne platforms.
		On the right, non-cooperative satellites may be used by unauthorized users for illegal or unregulated communications to bypass national regulations.
		To counter this threat, a signal monitoring model is deployed to detect anomalous signals and identify illicit transmissions.}
		\label{Fig1}
	\end{figure*}

	\section{DS2D Communications: Key Technologies and Application Scenarios} \label{section2}
	
	{\color{black}
	\subsection{System Architecture of DS2D Communications}
	The convergence of cooperative service delivery and non-cooperative security monitoring is critical for next-generation DS2D systems.
	In this paper, we propose a DS2D communication system comprising three architectural layers that enable both cooperative and non-cooperative functionalities: 
	(i) The physical layer encompasses the satellite constellation infrastructure, including LEO satellites equipped with multi-beam antennas, onboard computing units, and RF monitoring receivers. 
	This layer supports dual operational modes: cooperative mode for legitimate service delivery and monitoring mode for spectrum security governance.
	(ii) The network layer implements intelligent resource orchestration and security monitoring mechanisms.
	This layer manages user association, channel estimation, task offloading decisions, or unauthorized transmission identification.
	(iii) The application layer delivers end-to-end services to diverse user scenarios such as emergency communications in disaster areas, maritime connectivity, and aerial platform support.}
	
	\subsection{Key Technologies of DS2D Communications}
	
	\subsubsection{Satellite Beamforming and Beam Management}
	In DS2D communications, UEs are constrained by limited transmit power and antenna gain, making the link budget a critical factor for system availability.
	To address these limitations, satellites often employ multi-beam coverage and advanced beamforming techniques to achieve array gain, thereby supporting synchronization, access, and data transmission even under weak-signal conditions.
	For multi-beam management, the primary challenge lies in enabling UEs to perform beam discovery, measurement, and continuous tracking with minimal overhead.
	DS2D communications typically avoid exhaustive full-space scanning to identify the optimal beam, which can be computationally expensive and time-consuming. Instead, they prioritize lightweight beam selection strategies and controllable beam switching mechanisms.

	\subsubsection{Doppler Shift Estimation and Pre-Compensation}
	In satellite communications, Doppler shift presents a particularly acute challenge \cite{Lin2021Doppler}. The high relative velocity between LEO satellites and terrestrial UEs induces \textcolor{black}{a large and rapidly time-varying carrier-frequency offset}.
	If left unmitigated, such Doppler effects can severely disrupt subcarrier orthogonality in multi-carrier systems, leading to substantial inter-carrier interference (ICI) that degrades the reliability of synchronization, random access, and uplink transmission.
	To address this, DS2D systems typically adopt a lightweight closed-loop Doppler management strategy based on pre-compensation at the transmitter and residual tracking at the receiver. Specifically, the satellite pre-compensates for the predictable component of the downlink Doppler shift using precise orbital knowledge, while the UE estimates and corrects any residual offset during reception.

	\subsubsection{Multi-Satellite Coordination for DS2D}
	In multi-satellite systems, the DS2D communication shifts from a single satellite serving multiple users to a collaborative architecture enabled by overlapping coverage from multiple LEO satellites. When a user falls within the simultaneous coverage of several satellites, the system can exploit this redundancy to enable intelligent satellite selection, seamless handover, and even cooperative access, unlocking significant gains in performance and reliability.
	Such multi-satellite coordination enhances the effective data rate, reduces end-to-end latency, and improves resilience against signal blockage or temporary link outages caused by mobility or environmental obstructions.
	The true potential of multi-satellite DS2D lies not merely in increasing the number of visible satellites, but in enabling inter-satellite cooperation and scalable connectivity management \cite{Bakhsh2025Survey}.
	
	\subsubsection{Satellite-Terrestrial Cooperative Computing}
	As LEO satellites are increasingly equipped with onboard processors and memory units, satellite–terrestrial cooperative computing enabled by DS2D communications is emerging as a new paradigm for delivering edge intelligence services at truly global scale.
	By exploiting DS2D links, resource-constrained devices in remote, rural, or disaster regions can offload computation tasks not only to terrestrial servers but also directly to LEO satellites endowed with sufficient computing resources, or indirectly to terrestrial cloud infrastructure via satellite backhaul \cite{Peng2025I}.
	This hybrid architecture merges the high capacity and low latency of terrestrial networks with the pervasive coverage of NTNs, enabling seamless task offloading, distributed machine learning, and fast model inference.

	\subsection{Application Scenarios of DS2D Communications}
	As shown in Fig. \ref{Fig1}, \textcolor{black} {compared to conventional cellular networks, DS2D communications are unlocking new connectivity paradigms across a diverse range of challenging environments where terrestrial coverage is impractical, unavailable, or compromised.
	DS2D communications support a wide range of legitimate application scenarios (e.g., urban, highway, remote, disaster-response, airborne, and maritime connectivity) enabled by cooperative satellites, while simultaneously introducing security risks through potential misuse via non-cooperative satellites and unauthorized access channels.}
	\begin{itemize}
		\item
		\textbf{DS2D communications for remote regions and disaster areas:}
		In sparsely populated or geographically isolated areas, such as mountainous terrains, deserts, and polar regions, deploying terrestrial BSs is often \textcolor{black}{prohibitively expensive} or technically infeasible. Similarly, during natural disasters like earthquakes, floods, or wildfires, existing cellular infrastructure may be severely damaged or rendered inoperable. DS2D bridges this connectivity gap by providing immediate, wide-area communication capabilities without reliance on ground infrastructure. This ensures that affected populations and remote communities can maintain access to emergency alerts, voice messaging, and basic data services, thereby enhancing resilience and saving lives when terrestrial networks fail.
		\item
		\textbf{DS2D communications for maritime and oceanic scenarios:}
		Beyond coastal zones, maritime environments lie almost entirely outside the reach of terrestrial cellular networks. For commercial shipping, scientific research vessels, offshore energy platforms, and autonomous marine drones, sustained connectivity over open oceans is essential for navigation safety, regulatory compliance, environmental monitoring, and crew welfare. DS2D enables direct, global satellite links to standard maritime terminals, offering persistent coverage independent of proximity to shore.
		Meanwhile, DS2D provides robust backhaul and real-time data telemetry, transforming oceanic operations through seamless, end-to-end digital connectivity.
		\item
		\textbf{DS2D communications for aerial platforms:}
		For aerial platforms, such as aircraft and unmanned aerial vehicles (UAVs), terrestrial networks provide only limited and fragmented coverage during flight. DS2D overcomes this by establishing direct links between airborne user equipment and satellites.
		This enables a wide range of time-critical services, including UAV command-and-control, real-time surveillance, and remote operations.
	\end{itemize}

	\begin{figure*}[t]
		\centering
		\includegraphics[width=6.5 in]{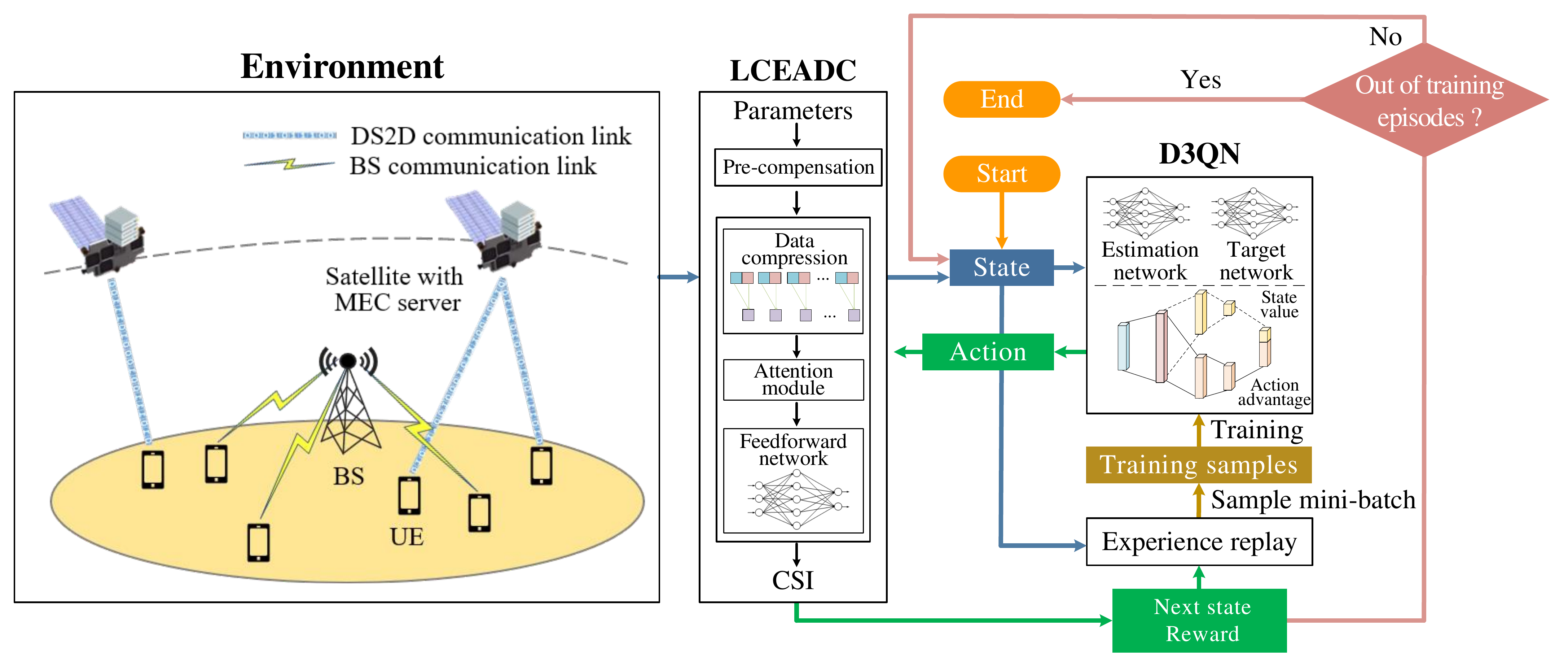}
		\caption{An illustration of the proposed algorithm for computing task offloading in cooperative DS2D communication scenarios}
		\label{Fig2}
	\end{figure*}

	\section{Computing Task Offloading via Cooperative DS2D Communications} \label{section3}
	
	LEO satellite networks have emerged as a promising solution for bridging the digital divide by providing ubiquitous communication and computing services to remote, rural, or underserved areas \cite{Peng2025I, Hassan2024Satellite}.
	However, the direct integration of user devices into LEO-based NTNs introduces significant operational challenges, as individual LEO satellites possess limited onboard computational resources and energy budgets.
	Without intelligent coordination, this surge in demand can degrade service quality and destabilize \textcolor{black}{the} DS2D system. Consequently, dynamic traffic steering is essential: users should be adaptively assigned to either satellite nodes or terrestrial BSs based on real-time channel conditions, computational load, and service requirements.
	This orchestration is further complicated by the highly dynamic LEO environment having severe Doppler shifts, fast-fading channels, and time-varying propagation delays \cite{He2026C}. These effects severely impair the accuracy and timeliness of CSI acquisition, rendering conventional, static resource allocation schemes ineffective.
	
	\subsection{System Model}
	As illustrated in Fig.~\ref{Fig2}, we consider an integrated satellite-ground network where each UE can flexibly select its access point, either a LEO satellite or a terrestrial BS, based on instantaneous channel conditions and service requirements.
	It is important to note that the terrestrial BS is not idle, it already serves a mix of ground users and IoT devices that lack satellite connectivity. If all UEs were to offload their tasks exclusively to the BS, despite its proximity and low propagation delay, the resulting traffic surge would overwhelm its computational capacity. This would trigger queuing delays, processing bottlenecks, and ultimately degrade overall response time, negating the benefits of terrestrial proximity.
	To address this trade-off, our objective is to minimize the average end-to-end latency across all UEs, which comprises both signal transmission delay and data processing time. By optimizing user–node associations, we aim to enhance quality of service (QoS), achieve system-wide load balancing, and improve resource utilization efficiency. Specifically, the decision variable is a binary association factor  $a_{u,n}(t) \in \{0,1\}$, indicating whether user $u$  is served by node $n$ (satellite or BS) in time slot $t$. We enforce the constraint that each user is assigned to exactly one service node per time slot, thereby preventing redundant resource consumption and ensuring efficient use of the hybrid space-ground infrastructure. \textcolor{black}{At each time slot, each UE generates a computation task, and the controller determines whether the task should be offloaded to a LEO satellite or to the terrestrial BS based on channel conditions, node loads, and service requirements. The UE then transmits the task data to the selected node for execution, after which the computation result is returned to the UE and the resulting latency is used to update the state and reward for the next decision epoch.}

	\subsection{Proposed Solution}	
	To address the joint challenges of dynamic channel estimation and intelligent user association in cooperative DS2D networks, we propose LCEADC-D3QN \cite{Lv2026Energy}, a novel DRL algorithm designed to minimize the average end-to-end latency across all UEs. The framework integrates two synergistic components:
	i)~a lightweight channel estimation with adaptive \textcolor{black}{Doppler} compensation (LCEADC) module for efficient and robust CSI acquisition under high-mobility LEO conditions, and
	ii)~a dueling double deep Q-network (D3QN) for dynamically optimizing user–node association decisions.
	Specifically, the LCEADC module tackles the difficulty of accurate channel estimation in fast-fading, Doppler-shifted satellite links. First, it leverages orbital ephemeris data from a spaceborne positioning system together with ground terminal location information to analytically compute and pre-compensate the expected Doppler frequency offset. Next, a compressed embedding mechanism reduces the dimensionality of raw time-series channel observations from 128 to 64 dimensions, while incorporating constellation geometric perception coding to preserve spatial context. To further lower computational overhead, we replace the standard self-attention mechanism with a lightweight linear attention module. Finally, a three-layer cascaded feedforward network, augmented with depthwise separable convolutions at each stage, refines the compressed features to produce an accurate estimate of the instantaneous CSI.
	\textcolor{black}{The LCEADC module outputs the Doppler-compensated instantaneous CSI rather than the final offloading decision. The estimated CSI is then incorporated into the state observed by the D3QN agent, together with node load information and user task demands. Based on this state,} the D3QN agent dynamically determines the optimal association for each user by maximizing long-term system performance.

	\begin{figure}[t]
		\centering
		\includegraphics[width=3.5 in]{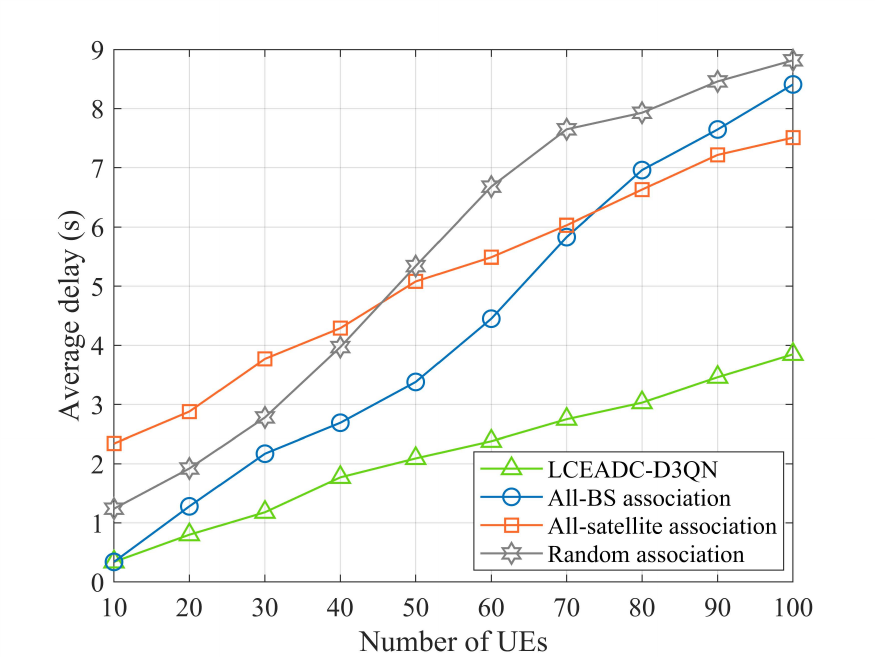}
		\caption{Simulation results for cooperative task offloading}
		\label{Fig3}
	\end{figure}
	
	\subsection{Simulation Results}
	To evaluate the performance of the proposed LCEADC-D3QN algorithm, we conduct simulations to compare average end-to-end latency across different user association strategies and \textcolor{black}{baseline algorithms}. As illustrated in Fig. \ref{Fig3}, the average latency grows monotonically with the number of UEs under all schemes, reflecting increasing system load.
	In the all-satellite association strategy, UEs experience high signal propagation delays due to the long satellite-ground path, compounded by limited onboard computational capacity, resulting in elevated processing latency. Conversely, the all-BS association strategy benefits from shorter terrestrial propagation delays but quickly becomes suboptimal as the BS becomes overloaded. This leads to severe queuing congestion and processing bottlenecks, ultimately degrading both data handling and response times.
	By contrast, our LCEADC-D3QN algorithm enables each UE to dynamically select the optimal access point, either a LEO satellite or the terrestrial BS, based on real-time channel conditions, node loads, and service demands.
	As a result, the proposed algorithm achieves significant reductions in both transmission and processing latency.
	Quantitatively, LCEADC-D3QN outperforms static association policies by up to 225\% in average latency reduction compared to the all-BS, all-satellite, and random association baselines.
	{\color{black}
	Our analysis reveals a fundamental trade-off in satellite-ground resource allocation. While terrestrial BSs offer lower propagation delay, their limited coverage and computational capacity create bottlenecks during peak demand. Conversely, LEO satellites provide global coverage but introduce higher propagation delay due to orbital altitude. The optimal strategy dynamically balances these factors based on real-time network conditions.}

	\section{Unauthorized Access Monitoring for Non-Cooperative DS2D Communications} \label{section4}
	
	\subsection{Signal Detection of Non-Cooperative DS2D}
	In an open-access DS2D environment, spectrum governance and \textcolor{black}{security monitoring} face unprecedented challenges.
	Specifically, the presence of unauthorized access introduces a complex coexistence scenario where \textcolor{black}{legitimate signals are mixed with interference from} non-cooperative satellite systems, posing a serious threat to national spectrum security and regulatory integrity.
	To counter this risk, it is imperative to develop real-time monitoring \textcolor{black}{capabilities} for unauthorized DS2D signals directly on satellites.
	However, conventional signal detection techniques exhibit critical limitations in the DS2D context.
	Energy detection, though computationally lightweight, is highly noise-sensitive and exhibits poor reliability at low SNRs, often yielding unstable performance due to threshold ambiguity. Cyclostationary feature detection, while more robust by exploiting inherent signal periodicities, requires long observation intervals and is significantly degraded by dynamic channel impairments, such as severe Doppler spread and time-varying fading.
	To overcome these drawbacks, recent work \cite{Nguyen2024WRIST} has proposed transforming received satellite signals into time–frequency spectrograms and applying object detection algorithms to identify signal regions in the image domain.
	Nevertheless, rectangular bounding boxes cannot accurately delineate the irregular time-frequency boundaries of signals affected by Doppler shifts and multipath effects.

	\begin{figure*}[t]
		\centering
		\includegraphics[width=6 in]{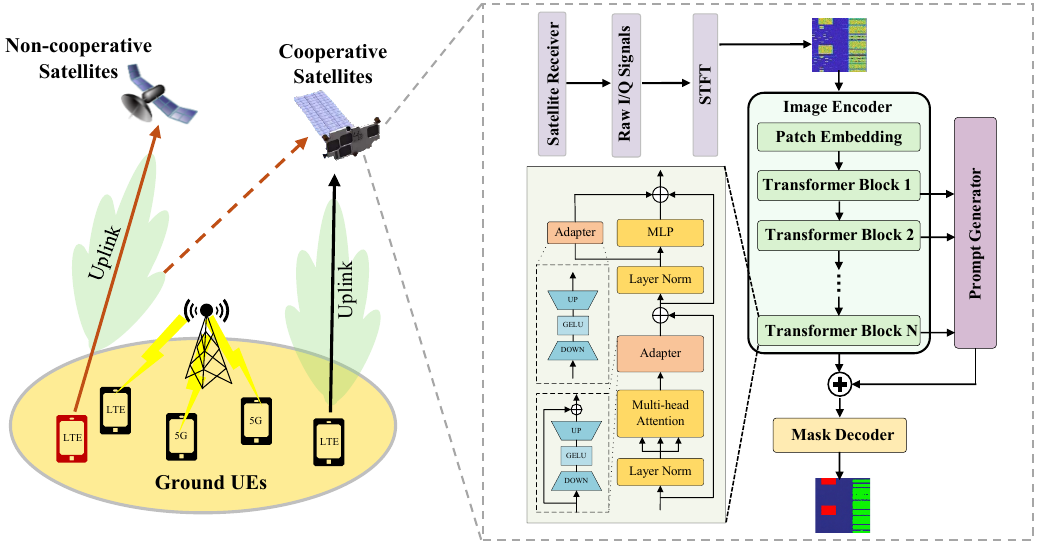}
		\caption{An illustration of the proposed algorithm for unauthorized signal detection in non-cooperative DS2D scenarios}
		\label{Fig4}
	\end{figure*}
	
	\begin{figure}[t]
		\centering
		\includegraphics[width=3.0 in]{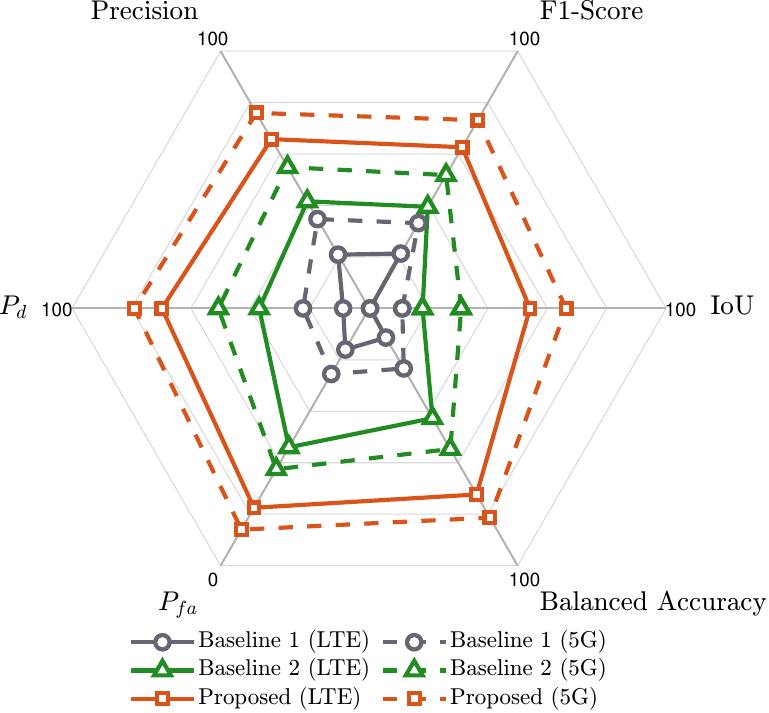}
		\caption{Simulation results for LTE and 5G NR signal detection}
		\label{Fig5}
	\end{figure}

	\subsubsection{System Model}	
	As illustrated in Fig. \ref{Fig4}, this section focuses on satellite-based signal monitoring in DS2D systems. The hybrid radio frequency signals received by the satellite are first converted into two-dimensional time-frequency spectrograms using the short-time \textcolor{black}{Fourier} transform (STFT), \textcolor{black}{thereby representing spectral dynamics as images.}
	This approach reframes the signal detection task as a multi-class semantic segmentation problem on time–frequency spectrograms. The model takes raw spectrograms as input and produces pixel-level semantic masks, where each pixel is classified into either a background category or one of several target categories corresponding precisely to the time–frequency occupancy of specific communication signals (e.g., LTE, 5G NR).
	Based on this pixel-level segmentation output, the system delivers two critical capabilities:
	i) Blind presence detection: It autonomously determines whether signals conforming to a known standard are present within a monitored frequency band, without requiring prior knowledge of timing, frequency, or modulation.
	ii) High-fidelity parameter extraction: By analyzing the geometric topology of the segmented signal regions, it directly estimates key physical-layer parameters, including center frequency, instantaneous bandwidth, and signal duration.
	These outputs not only significantly reduce the search space for subsequent signal acquisition, but also provide an intuitive and reliable basis for compliance verification through satellite-ground collaborative verification, facilitating the identification of illegal access or spectrum misuse.
	
	\subsubsection{Proposed Solution}
	To enable robust signal monitoring under \textcolor{black}{DS2D scenarios}, we propose a time-frequency spectrogram segmentation scheme based on SAM \cite{ye2025SAM}.
	The overall architecture is illustrated in Fig. \ref{Fig4}.
	Rather than performing full-parameter fine-tuning, which risks catastrophic forgetting and excessive resource consumption, we freeze the pre-trained ViT backbone to preserve its powerful feature extraction capabilities. To effectively bridge the significant domain discrepancy between natural images and spectrograms, we integrate lightweight adapter modules for efficient domain adaptation.
	To overcome SAM’s inherent dependence on manual prompts (e.g., user-provided points or boxes), we introduce an automatic prompt generation mechanism.
	Moreover, we redesign the mask decoder to support category-aware segmentation. By incorporating learnable category representations and optimizing feature upsampling pathways, the proposed model achieves end-to-end multi-class pixel-level segmentation, capable of distinguishing different communication standards (e.g., LTE, 5G NR) directly in the time–frequency domain.
	\textcolor{black}{As a result, the satellite can accomplish two tasks in a single forward pass:} i) determine the presence of target-standard signals, and ii)~perform precise time–frequency localization.
	This end-to-end capability seamlessly integrates rapid signal acquisition and network-side compliance verification, achieving efficient blind source monitoring in non-cooperative DS2D scenarios.
	
	\subsubsection{Simulation Results}
	To evaluate the effectiveness of the proposed signal detection scheme in non-cooperative DS2D scenarios, we select two representative terrestrial communication standards (e.g., LTE and 5G NR) as test cases.
	As shown in Fig. \ref{Fig5}, for LTE signals, the proposed method demonstrates clear superiority over baseline approaches in both localization precision and detection reliability.
	Specifically, compared to Baseline 2, it achieves a 21.7\% improvement in intersection over union (IoU) and a 13.9\% gain in F1-score, while reducing the false alarm rate from 9.2\% to 4.5\%.
	For 5G NR signals, all methods exhibit \textcolor{black}{better performance than that obtained for LTE signals}, reflecting the inherently richer time–frequency structure and greater feature separability of 5G NR waveforms.
	Similarly, our proposed solution \textcolor{black}{consistently outperforms the baselines} across all key metrics, including IoU, F1-score, and false alarm rate.
	Collectively, these results confirm that the proposed solution significantly enhances pixel-level segmentation accuracy and false alarm suppression for satellite spectrum monitoring and unauthorized access identification in open-access DS2D networks.

	\begin{figure*}[t]
		\centering
		\includegraphics[width=6.5 in]{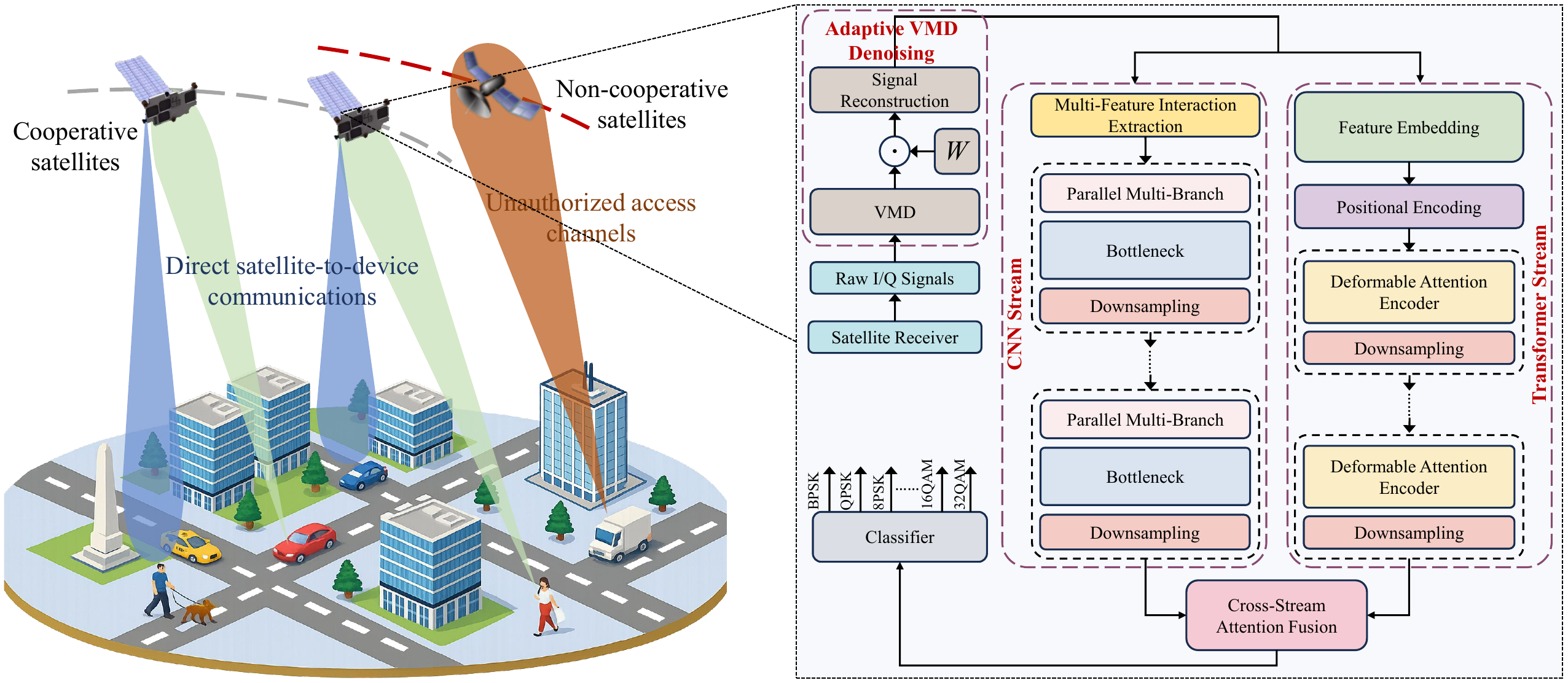}
		\caption{An illustration of the proposed algorithm for signal modulation classification in non-cooperative DS2D scenarios}
		\label{Fig6}
	\end{figure*}
	
	\begin{figure}[t]
		\centering
		\includegraphics[width=3.0 in]{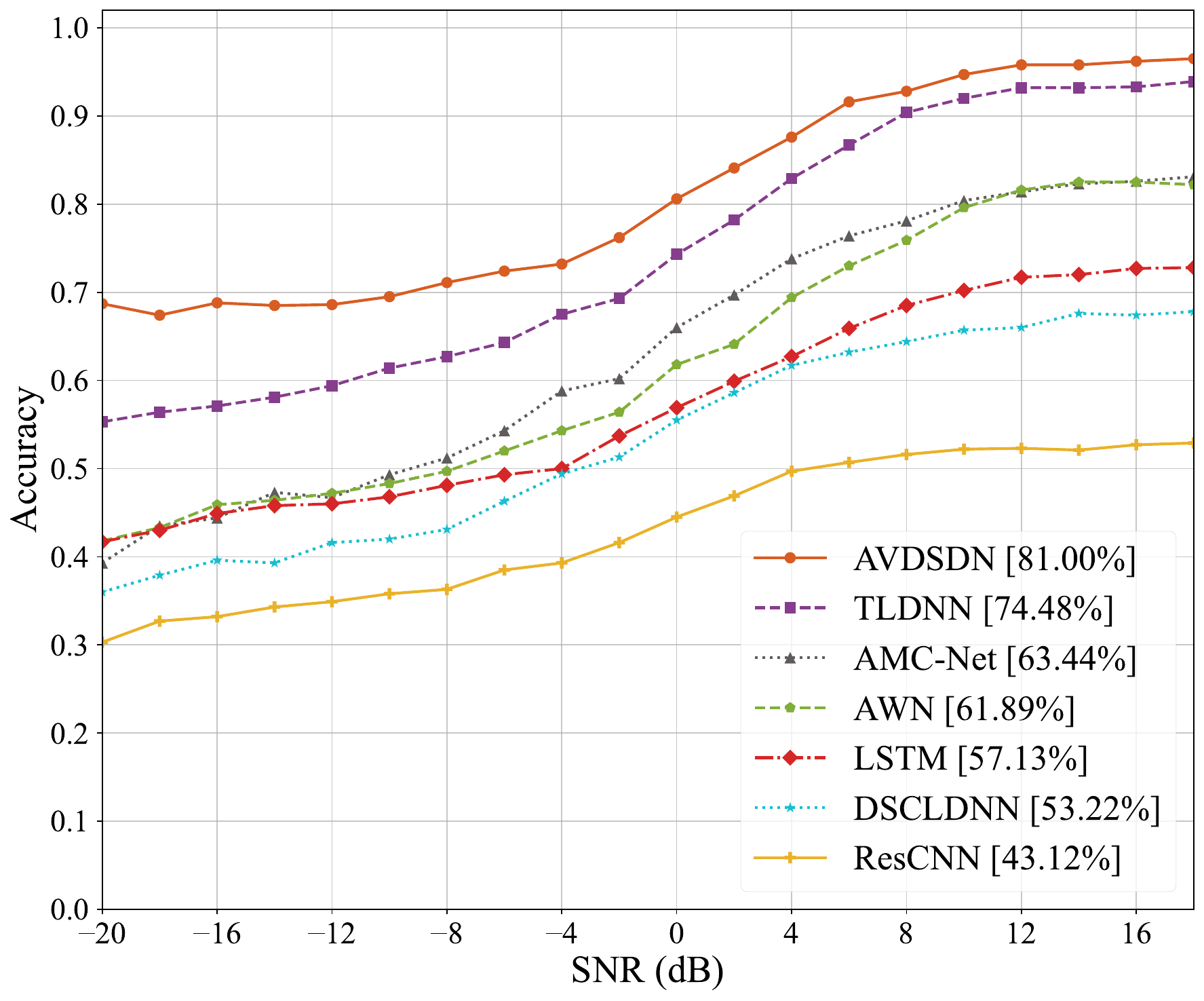}
		\caption{AMC accuracy versus SNR in non-cooperative satellite scenarios}
		\label{Fig7}
	\end{figure}
	
	\subsection{Onboard AMC for Non-Cooperative DS2D Access}	
	To safeguard spectrum security and national sovereignty, it is important for cooperative satellites to autonomously monitor and identify unauthorized access between illegal users and foreign non-cooperative satellites.
	Onboard AMC offers a promising solution by enabling real-time signal identification directly on satellite platforms.
	However, this solution is challenged by two key constraints: the stringent limitations on onboard computational resources and the dynamic nature of satellite channels.
	Traditional AMC approaches are ill-suited to the satellite network due to their high complexity or sensitivity to channel distortions. While deep learning (DL)-based methods have emerged as the dominant paradigm for AMC, naively transferring terrestrial DL models to the satellite context often fails to account for the non-stationary, time-varying, and bursty characteristics of intercepted modulated signals in LEO scenarios.
	Compounding these issues, monitoring satellites typically intercept signals through antenna sidelobes, resulting in extremely low SNRs. Conventional denoising techniques, designed primarily for stationary signals, perform poorly on the transient and spectrally sparse waveforms typical of DS2D transmissions.
	To address these challenges, we advocate for a robust AMC algorithm for low-SNR non-cooperative DS2D access monitoring by synergistically fusing spatial and temporal features.
	
	\subsubsection{System Model}
	As illustrated in Fig. \ref{Fig6}, we consider an unauthorized access monitoring system targeting non-cooperative satellite links, wherein a cooperative LEO satellite is tasked with monitoring illicit communications between ground terminals and unauthorized (non-cooperative) satellites.
	Since the monitoring satellite typically lies outside the main lobe of the terminal's transmit antenna pattern, it receives signals primarily through antenna sidelobe leakage, resulting in inherently low SNRs.
	To accurately reflect the physical propagation environment, \textcolor{black}{the received signal model incorporates physical-layer impairments} such as Doppler frequency shifts, time-varying propagation delays, path loss, and multipath effects in dynamic channel environments.
	Then, we formulate the AMC task as a multi-class supervised learning problem under non-cooperative conditions.
	The objective is to establish a nonlinear mapping from raw or preprocessed signal observations to modulation class labels, enabling precise identification of unknown DS2D signal types.

	\subsubsection{Proposed Solution}
	\textcolor{black}{To achieve reliable modulation recognition in non-cooperative DS2D scenarios, we propose an adaptive variational mode decomposition (VMD) denoising-based synergistic dual-stream network (AVDSDN) \cite{Zheng2026Synergistic}, which comprises three core components: i) an adaptive VMD denoising module, ii) a dual-stream feature extraction backbone, and iii) a cross-stream attention fusion module, designed to jointly optimize denoising and classification processes. Different from conventional serial or homogeneous parallel AMC algorithms, AVDSDN decouples spatial and temporal feature extraction, leveraging a dual-stream architecture to capture discriminative features from different signal dimensions and coupling them toward the final AMC objective. This design improves the robustness of the model against low-SNR conditions and time-varying modulated signals.}
	Specifically, the adaptive denoising module uses VMD to decompose the signal and performs task-driven weighted reconstruction, enhancing modulation-relevant features while suppressing noise.
	The cleaned signal is then processed in parallel by a dual-stream feature extraction module: the CNN stream captures multi-scale local spectral features with a lightweight backbone and soft thresholding to filter residual noise, while the Transformer stream employs deformable attention to dynamically focus on key temporal regions and integrates a convolutional feed-forward network for efficient long-range modeling.
	Finally, a cross-stream attention fusion module combines spatial and channel-wise information from both streams, producing a discriminative representation that significantly improves AMC accuracy under low-SNR conditions.
	
	\subsubsection{Simulation Results}
	In the experiments, we consider 10 modulation types: BPSK, QPSK, 8PSK, 16PSK, 4PAM, 8PAM, 32QAM, 64QAM, 128QAM, and 256QAM.
	As shown in Fig. \ref{Fig7}, the proposed AVDSDN outperforms state-of-the-art methods (e.g., AMC-Net, TLDNN, and AWN) across the entire SNR range.	
	Notably, AVDSDN demonstrates exceptional robustness under low-SNR conditions, achieving 70.44\% average recognition accuracy in the -20 dB to 0 dB SNR range, which is higher than the 61.15\% achieved by the TLDNN.
	\textcolor{black}{Moreover, by employing INT8 post-training quantization, AVDSDN achieves low-latency inference with negligible accuracy loss. Combined with the GPU-accelerated VMD denoising stage, the end-to-end processing latency is approximately 5-6 ms per signal segment, demonstrating its practicality for deployment on resource-constrained satellite platforms.}

	\section{Future Research Directions} \label{section5}
	To realize multi-functional space-ground integrated networks, we outline several promising research directions that can further advance satellite-enabled ubiquitous connectivity.
	
	\subsection{Integrated Satellite Communication, Sensing, Positioning, and Navigation}
	The convergence of communication and sensing in 6G opens new avenues for joint space-ground service delivery. Future DS2D systems may support unified waveform designs that simultaneously enable data transmission, environmental sensing, high-accuracy positioning, and navigation, particularly in GPS-denied or remote regions. By leveraging LEO satellites’ global coverage and mobility, such integrated systems could provide resilient situational awareness for autonomous vehicles, disaster response, and maritime operations, while sharing spectral and hardware resources to improve energy efficiency.
	
	\subsection{FL for Distributed Space-Ground Intelligence}
	With growing computational capabilities on satellites and edge devices, distributed machine learning paradigms \textcolor{black}{are becoming essential} to preserve privacy and reduce backhaul load. Federated learning (FL) offers a promising framework wherein ground terminals collaboratively train global models using local data, while only exchanging model parameters or knowledge via satellite links. Future research should address the challenges of FL in DS2D contexts, including intermittent connectivity, heterogeneous device capabilities, and straggler mitigation under orbital dynamics. Onboard satellite intelligence could further act as an FL orchestrator, enabling adaptive model aggregation across continental-scale user populations.
		
	\subsection{Software-Defined Reconfigurable Satellite Networks}
	To accelerate innovation and reduce lifecycle costs, future LEO satellites must embrace software-defined networking principles. A critical enabler is in-orbit software reconfiguration, which allows satellites to update communication protocols, modulation schemes, or signal processing algorithms via remote firmware upgrades, eliminating the need to launch replacement hardware. Coupled with network function virtualization, this capability would permit dynamic instantiation of virtualized network services (e.g., emergency alert broadcasting, IoT data relay, or spectrum monitoring) on shared satellite platforms. Research should explore robust over-the-air update mechanisms, fault-tolerant execution environments, and standardized interfaces for cross-vendor interoperability in heterogeneous mega-constellations.
	
	\subsection{Lightweight Security Protocols for DS2D Communications}
	Security remains a paramount concern in DS2D systems due to the broadcast nature of satellite transmissions and the limited computational resources of handsets. Conventional cryptographic protocols incur prohibitive latency and energy overhead, rendering them unsuitable for low-power IoT or emergency terminals. Future efforts should focus on designing lightweight cryptographic primitives, such as post-quantum secure symmetric ciphers and compact digital signatures, that offer strong security guarantees with minimal footprint. Complementary physical-layer security techniques, including directional beamforming, artificial noise injection, and channel-based key generation, can further enhance confidentiality without relying solely on upper-layer encryption. Additionally, decentralized trust management via blockchain or lightweight consensus mechanisms may strengthen authentication in satellite networks.
	
	\subsection{Theoretical Foundations of Satellite-Ground Integrated Networks}
	First, the information-theoretic limits of DS2D communications under high-mobility and rapidly time-varying channel conditions are poorly understood, as existing capacity bounds fail to capture the unique Doppler spreads, intermittent connectivity, and non-stationary fading characteristics inherent to LEO environments.
	Second, there exists a fundamental lack of analytical frameworks characterizing the trade-offs between latency, reliability, and energy efficiency in satellite edge computing systems having the ultra-long propagation delays and intermittent ground-satellite links.
	Collectively, these gaps impede the design of next-generation integrated satellite-terrestrial networks that can guarantee quality-of-service across diverse application scenarios.
	
	\section{Conclusion} \label{section6}
	This article presented a comprehensive framework for DS2D communications that addresses both cooperative service delivery and non-cooperative security monitoring challenges.
	For cooperative task offloading, we proposed a novel D3QN algorithm to estimate channel conditions and dynamically orchestrate user association for latency minimization.
	For non-cooperative access monitoring, we proposed Transformer-based \textcolor{black}{models to effectively detect and classify unauthorized transmissions}, even under challenging low-SNR conditions.
	Simulation results showed that the cooperative task offloading algorithm achieved up to 225\% latency reduction compared to static association policies, while the non-cooperative monitoring model maintained 90.5\% detection probability for unauthorized access attempts.
	Looking forward, as LEO constellations continue to expand, future work should focus on developing advanced resource orchestration algorithms, cross-domain learning techniques, and standardized interface specifications to enable seamless integration across heterogeneous satellite networks. Ultimately, the successful realization of unified DS2D frameworks will require collaborative efforts from industry, academia, and regulatory bodies to establish technical standards and operational best practices that ensure secure, efficient, and globally accessible satellite communications.

\end{document}